\documentstyle[aps,prb,graphics,rotating,epsfig,multicol]{revtex}

\def\bs#1{\bbox{#1}}
\def\ve#1{\bs{\mathbf{#1}}}
\newcommand{\be}{\begin{eqnarray}}
\newcommand{\ee}{\end{eqnarray}}

\begin{document}

\draft \title{Coupled Dipole Method Determination of the
Electromagnetic Force on a Particle over a Flat Dielectric Substrate}
\author{P. C. Chaumet and M. Nieto-Vesperinas}

\address{Instituto de Ciencia de Materiales de Madrid, Consejo
Superior de investigaciones Cientificas, Campus de Cantoblanco Madrid
28049, Spain}

\maketitle

\begin{abstract}

We present a theory to compute the force due to light upon a particle
on a dielectric plane by the Coupled Dipole Method (CDM). We show
that, with this procedure, two equivalent ways of analysis are
possible, both based on Maxwell's stress tensor. The interest in using
this method is that the nature and size or shape of the object, can be
arbitrary. Even more, the presence of a substrate can be
incorporated. To validate our theory, we present an analytical
expression of the force due to the light acting on a particle either
in presence, or not, of a surface. The plane wave illuminating the
sphere can be either propagating or evanescent. Both two and three
dimensional calculations are studied.

\end{abstract}


%\twocolumn
\begin{multicols}{2}
\section{Introduction}

The demonstration of mechanically acting upon small particles with
radiation pressure was done by Ashkin and
coworkers~\cite{ashkin69,ashkin70}. A consequence of these works was
the invention of the optical tweezer for non destructive manipulation
of suspended particles~\cite{ashkin86} or molecules and other
biological objects~\cite{ashkin87,block89,ashkin97}. Recently, these
studies have been extended to the nanometer
scale~\cite{novotny,tanaka,renn,svoboda,sugiura,omori}, and multiple
particle configurations based on optical binding have been
studied~\cite{burns,gu,anto1,anto2,bayer}. Also, the effect of
evanescent waves created by total internal reflection on a dielectric
surface on which particles are deposited was studied in
Ref.~[\ref{kawata}]. However, the only theoretical interpretation of
such system is given in Refs.~[\ref{almaas}-\ref{lester}]. In
Ref.~[\ref{almaas}] no multiple interaction of the light between the
particles and the dielectric surface was taken into account. On the
other hand, in Ref.~[\ref{lester}] a multiple scattering numerical
method was put forward limited to a 2-D configuration.

It is worth remarking here that several previous theoretical works on
optical forces usually employ approximations depending on the radius
of the particle; if the particle is small it has been usual to split
the force into three parts: the gradient, scattering, and absorbing
forces~\cite{visscher}. However, a rigorous and exact calculation
requires the use of Maxwell's stress tensor. We shall use it in this
paper. Some work has been done in free space~\cite{novotny,barton},
or for a spherical particle over a dielectric surface illuminated by a
Gaussian beam.~\cite{chang}

We shall present, therefore, a detailed theoretical analysis in three
dimensions of how the optical force is built on the multiple
interaction of light with the particle and the dielectric surface.
This will be done whatever its size, shape, or permittivity. To this
end, we shall make use of the Coupled Dipole Method (CDM), whose
validity was studied in detail in Ref.~[\ref{chaumet}].

In Section~\ref{sectioncdm} we present the CDM, and two possibilities
that arise with this method to compute the force by means of Maxwell's
stress tensor. Concerning the first one, in Section~\ref{CDMA} we use
directly Maxwell's stress tensor and perform the surface
integrations. As regards the second one, we present in
Section~\ref{CDMB} the dipole approximation on each subunit of
discretization for the numerical calculations. Since however these
methods are somewhat cumbersome from a numerical point of view, we
have introduced in Section~\ref{CDMparticle} an analytical calculation
for the force due to the light on a small particle in the presence of
the surface. Results are illustrated in three dimensions in
Section~\ref{dipapps} (a little sphere) and in two dimensions in
Section~\ref{dipappc} (a small cylinder). In Section~\ref{result} we
compute the force with the CDM, and we validate these calculations on
electrically small particles by means of the analytical solution
presented in Section~\ref{CDMparticle}. After this validation of the
CDM on little particles, we present in Section~\ref{bigsphere}
calculations on larger particles.

\section{Electromagnetic force computed with the Coupled Dipole Method}\label{sectioncdm}

The Coupled Dipole Method (CDM) was introduced by Purcell and
Pennypacker in 1973 for studying the scattering of light by
non-spherical dielectric grains in free space.\cite{purcell} This
system is represented by a cubic array of $N$ polarizable subunits.
The electric field $\ve{E}(\ve{r}_i,\omega)$ at each subunit position
$\ve{r}_i$ can be expressed as: \be \label{dipi}
\ve{E}(\ve{r}_i,\omega) & = & \ve{E}_0(\ve{r}_i,\omega) +
\sum_{j=1}^{N} [ \ve{S}(\ve{r}_i,\ve{r}_j,\omega)\\\nonumber & + &
\ve{T}(\ve{r}_i,\ve{r}_j,\omega)] \alpha_j(\omega)
\ve{E}(\ve{r}_j,\omega). \ee where $\ve{E}_0(\ve{r}_i,\omega)$ is the
field at the position $\ve{r}_i$ in the absence of the scattering
object, $\ve{T}$ is the field susceptibility associated to the free
space~\cite{jackson:75} (with $\ve{T}(\ve{r}_i,\ve{r}_i,\omega)=0$ to
avoid the diagonal case), and $\ve{S}$ represents the field
susceptibility associated with the surface in front of which the
particle is placed (see Fig.~1). The derivation of the field
susceptibility of the surface is extensively developed in
Refs.~[\onlinecite{agarwal,rahmani}]. $\alpha_j(\omega)$, the
polarizability of the subunit $j$, is expressed as: \be\label{cmreact}
\alpha_j(\omega)=\alpha_j^{0}(\omega)/\left[1-(2/3)i k_0^3
\alpha_j^{0}(\omega)\right]\ee where $k_0=|\ve{k}_0|=\omega/c$
($\ve{k}_0$ being the incident wavevector of the electromagnetic field
in vacuum) and $\alpha_j^{0}(\omega)$ is given by the
Clausius-Mossotti relation~:

\be\label{cmcdm} \alpha_j^{0}(\omega) & = & \frac{3d^3}{4\pi}
\frac{\varepsilon(\omega)-1}{\varepsilon(\omega)+2}.  \ee In
Eq.~(\ref{cmcdm}) $d$ is the spacing of lattice discretization and
$\varepsilon(\omega)$ stands for the relative permittivity of the
object. Let us remark that the polarizability is expressed according
to Eq.~(\ref{cmreact}) as defined by Draine~\cite{draine}. The term
$(2/3)i k_0^3 \alpha_j^{0}(\omega)$ is the radiative reaction term,
necessary for the optical theorem to be satisfied and for a correct
calculation of forces via the CDM.~\cite{opl}

\begin{figure}[H]
\begin{center}
\resizebox{60mm}{!}{\input{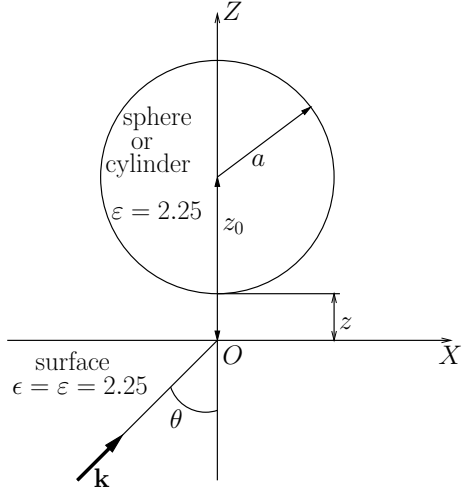}}
\end{center}
\caption{Geometry of the configuration considered in this paper:
sphere, or cylinder, of radius $a$ on a dielectric flat surface. The
relative permittivity is $\varepsilon=2.25$ both for the sphere (or
the cylinder) and the surface. The wavelength used is
$\lambda=632.8$nm in vacuum and the incident wave vector $\ve{k}$ is
in the $XZ$ plane.}
\end{figure}

Once the values of $\ve{E}(\ve{r}_i,\omega)$ are obtained by solving
the linear system, Eq.~(\ref{dipi}), (whose size is $3N\times3N$), it
is easy to compute the field at an arbitrary position $\ve{r}$: \be
\label{dipr} \ve{E}(\ve{r},\omega) & = & \ve{E}_0(\ve{r},\omega) +
\sum_{j=1}^{N} [ \ve{S}(\ve{r},\ve{r}_j,\omega)\\\nonumber & + &
\ve{T}(\ve{r},\ve{r}_j,\omega)] \alpha_j(\omega)
\ve{E}(\ve{r}_j,\omega). \ee The computation of the force also
requires the magnetic field radiated by the scattering object. We
obtain it through Faraday's equation,
$\ve{H}(\ve{r},\omega)=c/(i\omega)
\ve{\nabla}\times\ve{E}(\ve{r},\omega)$

\subsection{Force computed with Maxwell's stress tensor}\label{CDMA}

The force $\ve{F}$ on an object due to the electromagnetic
field~\cite{force} is computed from Maxwell's stress
tensor:~\cite{stratton}

\be \label{forcemaxwell}\ve{F} &=& 1/(8\pi)\Re e\bigg[ \int_S\big[
(\ve{E}(\ve{r},\omega).\ve{n})\ve{E}^{*}(\ve{r},\omega)\\\nonumber &+&
(\ve{H}(\ve{r},\omega).\ve{n})\ve{H}^{*}(\ve{r},\omega)\\
\nonumber & - & 1/2( |\ve{E}(\ve{r},\omega)|^{2}+
|\ve{H}(\ve{r},\omega)|^{2})\ve{n}\big] d\ve{r}\bigg], \ee where $S$
is a surface enclosing the object, $\ve{n}$ is the local outward unit
normal, $*$ denotes the complex conjugate, and $\Re e$ represents the
real part of a complex number. Let us notice that
Eq.~(\ref{forcemaxwell}) is written in CGS units for an object in
vacuum, and so will be given all forces presented in
Section~\ref{result}. To apply Eq.~(\ref{forcemaxwell}) with the CDM,
we must first solve Eq.~(\ref{dipi}) to obtain
$\ve{E}(\ve{r}_i,\omega)$ at each dipole position, and then, through
Eq.~(\ref{dipr}) and Faraday equation, the electromagnetic field is
computed at any position $\ve{r}$ of $S$. This enables us to
numerically perform the two dimensional quadrature involved in
Eq.~(\ref{forcemaxwell}).

\subsection{Force determined via the dipolar approximation}\label{CDMB}

Let us consider a small spherical particle with a radius smaller than
the wavelength. Then the $u$-component of the force can be written in
the dipole approximation:~\cite{gordon,opl} \be\label{forcec}
F_u(\ve{r}_0) & = & (1/2)\Re e\sum_{v=1}^{3}\left(p_v(\ve{r}_0,\omega)
\frac{\partial E_v^*(\ve{r}_0,\omega)}{\partial u}\right),\\\nonumber
& & \text{ ($u$=1, 2, 3)} \ee where $\ve{r}_0$ is the position of the
center of the sphere and $u$ and $v$ stand for the components along
either $x,y$ or $z$. We discretize the object into $N$ small dipoles
$\ve{p}_i(\ve{r},\omega)$ ($i$=1,...,$N$) so that it is possible to
compute the force on each dipole from Eq.~(\ref{forcec}). Hence, to
obtain the total force on the particle it suffices to sum the
contributions $\ve{F}(\ve{r}_i)$ from of all the dipoles. To use this
method it is necessary to know $\displaystyle{\frac{\partial
E_v(\ve{r}_i,\omega)}{\partial u}}$ at each discretization subunit. On
performing the derivative of Eq.~(\ref{dipi}) we obtain:

\be & & \left(\frac{\partial
  \ve{E}(\ve{r},\omega)}{\partial\ve{r}}\right)_{\ve{r}=
\ve{r}_i}  =  \left(\frac{\partial \ve{E}_0(\ve{r},\omega)}{\partial
\ve{r}}\right) _{\ve{r}=\ve{r}_i}\\\nonumber & + & \sum_{j=1}^{N}
\left(\frac{\partial}{\partial\ve{r}}[
\ve{S}(\ve{r},\ve{r}_j,\omega)+\ve{T}(\ve{r},\ve{r}_j,\omega)]
\right)_{\ve{r}=\ve{r}_i}\alpha_j(\omega) \ve{E}(\ve{r}_j,\omega).
\ee

Thus, the derivative of the field at $\ve{r}_i$, requires that of
$\ve{E}_0(\ve{r}_i,\omega)$ and that of the field susceptibility both
in free space and in the presence of the surface for all pairs
$(\ve{r}_i,\ve{r}_j)$. Hence we now have two tensors with 27
components each. It is important to notice that the derivative of the
field at $\ve{r}_i$ has been directly computed from just the field at
this position $\ve{r}_i$, so it is not computed in a self-consistent
manner. To have the required self-consistence for the derivative, it
is necessary to perform in Eq.~(\ref{dipi}) a multipole expansion up
to second order. Then, this equation must be written up to the
quadrupole order after taking its derivative. As a result, we obtain a
linear system, whose unknowns are both the electric field and its
derivative. The disadvantage of this method is that the size of the
linear system increases up to $12N\times12N$ and requires the
computation of the second derivative of the field susceptibility (81
components). More information about the CDM by using the multipole
expansion can be found in Ref.~[\ref{chaumet}].

In what follows, we shall denote CDM-A the force computed directly
from Maxwell's stress tensor Eq.~(\ref{forcemaxwell}) and CDM-B the
force obtained on using the field derivative Eq.~(\ref{forcec}). The
advantages of these two methods is that they are not restricted to a
particular shape of the object to be discretized. Furthermore, this
object can be inhomogeneous, metallic, or in a complex system whenever
it is possible to compute its field susceptibility.

\section{Force on a dipolar particle}\label{CDMparticle}

\subsection{The three dimensional case: a sphere}\label{dipapps}

Eq.~(\ref{dipi}) with $N=1$, taking the surface into account, gives
for the field at the position $\ve{r}_0=(x_0,y_0,z_0)$ of the sphere
of a radius $a$: \be
\label{forcedips}\ve{E}(\ve{r}_0,\omega)
=\left[\ve{I}-\alpha(\omega)\ve{S}(\ve{r}_0,\ve{r}_0,\omega)\right]
^{-1}\ve{E}_0(\ve{r}_0,\omega), \ee where $\ve{I}$ is the unit tensor,
and $\alpha(\omega)$ the polarizability of the sphere according to
Eq.~(\ref{cmreact}) with
$\alpha_0(\omega)=a^3(\varepsilon(\omega)-1)/(\varepsilon(\omega)+2)$. We
notice that $\ve{S}$ is purely diagonal and depends only on the
distance $z_0$ between the center of the sphere and the surface (see
Fig.~1). We also assume that the sphere is near the surface and,
hence, the field susceptibility of the surface can be used in the
static approximation ($k_0=0$, we shall discuss the validity of this
approximation in Section~\ref{result}). Therefore, the components of
this tensor become $S_{xx}=S_{yy}=-\Delta/(8 z_0^{3})$, and
$S_{zz}=-\Delta/(4 z_0^{3})$, with $\Delta=(1-\epsilon)/(1+\epsilon)$
representing the Fresnel coefficient of the surface. Since we consider
the object in the presence of a surface with a real relative
permittivity, $\Delta$ is real. As shown by Fig.~1, the light incident
wave vector $\ve{k}_0$lies in the $XZ$ plane. Therefore, there is no
force in the $Y$-direction. On using Eqs.~(\ref{forcec})
and~(\ref{forcedips}), and assuming the incident field ${\ve{E}_0}$
above the surface to be a plane wave either propagating or evanescent,
depending on the illumination angle $\theta$, the components of the
force on the sphere can be written as: \be\label{force3dpa} F_x & =
&\frac{\Re e}{2}\left[4\alpha z_0^3(i k_x)^{*}
\left(\frac{2|{E_0}_x|^{2}}{8z_0^{3}+\alpha\Delta}+
\frac{|{E_0}_z|^{2}}{4z_0^{3}+\alpha\Delta}\right)\right].\\
\label{force3dpb} F_z & = &
|{E_0}_x|^{2}\frac{\Re e}{2}\left( \frac{8z_0^3\alpha(i k_z)^{*}}
{8z_0^{3}+\alpha\Delta}+\frac{12z_0^2|\alpha|^2\Delta}
{|8z_0^{3}+\alpha\Delta|^{2}}\right)\\\nonumber & + & |{E_0}_z|^{2}\frac{\Re
e}{2}\left( \frac{4z_0^3\alpha(i k_z)^{*}}
{4z_0^{3}+\alpha\Delta}+\frac{6z_0^2|\alpha|^2\Delta}
{|4z_0^{3}+\alpha\Delta|^{2}}\right).\ee for $p$-polarization and
\be\label{force3dsa} F_x & = & |{E_0}_y|^{2} \frac{\Re e}{2}\left[
\frac{8z_0^3\alpha(ik_x)^*}{8z_0^{3}+\alpha\Delta}\right]\\
\label{force3dsb} F_z & = & |{E_0}_y|^{2}\frac{\Re e}{2}
\left( \frac{8z_0^3\alpha(i k_z)^{*}}
{8z_0^{3}+\alpha\Delta}+\frac{12z_0^2|\alpha|^2\Delta}
{|8z_0^{3}+\alpha\Delta|^{2}}\right).\ee for $s$-polarization.  We see
that the advantage of working with the static approximation is that an
analytic form of the force is obtained. To see the effect of the
incident field only (i.e., without interaction with the surface), we
can put $z_0\rightarrow\infty$ or $\Delta=0$ in
Eqs.~(\ref{force3dpa})-(\ref{force3dsb}). The forces are then
expressed as: \be\label{forceairx} F_x&=&|{E_0}|^{2}\frac{\Re
e}{2}\left( \alpha(ik_x)^{*}\right),\\
\label{forceairz}
F_z&=&|{E_0}|^{2}\frac{\Re e}{2}\left( \alpha(ik_z)^{*}\right), \ee
with $|{E_0}|^{2}=|{E_0}_y|^{2}$ for $s$-polarization and
$|{E_0}|^{2}=|{E_0}_x|^{2}+|{E_0}_z|^{2}$ for $p$-polarization.
Eqs.~(\ref{forceairx})-(\ref{forceairz}) show a spherical symmetry,
and hence the results both in $p$ and $s$-polarization are the same.

If we look at Fig.~1, we see that the incident field above the surface
always has $k_x$ real, but $k_z$ can be either real (propagating wave)
or imaginary (evanescent wave when $\theta>\theta_c$ where $\theta_c$
is the critical angle defined as
$\sqrt{\epsilon}\sin\theta_c=1$). Hence, all forces in the
$X$-direction have the form $A\Re e( \alpha (ik_x)^*)$ where $A$ is
always a positive number. In using Eq.~(\ref{cmreact}) we find that
$\Re e( \alpha (ik_x)^*)\simeq(2/3)\alpha_0^2k_0^3k_x$ (we have
assumed that $(4/9)k_0^6\alpha_0^2\ll 1$, in fact this expression is
about $6.6\times 10^{-7}$ for $a=10$nm, $\lambda=632.8$nm, and
$\varepsilon=2.25$, thus this approximation is perfectly
valid). Hence, whatever the field, either propagating or evanescent,
and the system either in presence of a surface or in free space, the
force in the $X$-direction is always along the incident field.

From Eq.~(\ref{forceairz}) and from the discussion above, it is easy
to see that in the absence of interface the force is positive for a
propagating incident wave ($k_z$ real). In the case of an evanescent
incident wave, $k_z=i\gamma$ with $\gamma>0$, and hence the force
becomes $F_z=-\gamma\alpha_0|E_0|^2/2$; namely the sphere is attracted
towards the higher intensity field. Concerning the force along the
$Z$-direction, its sign will depend on the nature of the field and the
interaction of the sphere with the surface. We shall discuss this in
Section~\ref{resultsphere}.

\subsection{The two dimensional case: a cylinder}\label{dipappc}

For a cylinder with its axis at $(x_0,z_0)$, parallel to the $Y$-axis
(Fig.~1), the electric field at its center is obtained by an equation
similar to Eq.~(\ref{forcedips}), but with a different
polarizability. With the help of Refs.~[\ref{lakhtakia}]
and~[{\ref{yaghjian}] we write this polarizability: \be
\alpha_1(\omega)& = & \frac{\alpha_1^{0}(\omega)}
{1-ik_0^{2}\pi\alpha_1^{0}(\omega)/2},\text{\hskip2mm with\hskip2mm}
\alpha_1^{0}(\omega)=\frac{\varepsilon(\omega)-1}
{\varepsilon(\omega)+1}\frac{a^{2}}{2}.\\ \alpha_2(\omega)& = &
\frac{\alpha_2^{0}(\omega)}{1-ik_0^{2}\pi\alpha_2^{0}(\omega)},
\text{\hskip2mm with \hskip2mm}
\alpha_2^{0}(\omega)=(\varepsilon(\omega)-1)\frac{a^{2}}{4}.\ee The
subscripts $i=1$, and 2 correspond to the field either perpendicular
or parallel to the axis of the cylinder, respectively. The field
susceptibility of the surface in the two dimensional case is given in
Ref.~[\ref{jjg2}] for $s$-polarization and in Ref.~[\ref{jjg3}] for
$p$-polarization. Since we address a cylinder with a small radius $a$
and near the surface, we use the static approximation, then
$S_{xx}=S_{zz}=-\Delta/(2z_0^{2})$ and $S_{yy}=0$. In the same way as
seen before, the force is written as: \be F_x&=&|{E_0}|^{2}\frac{\Re
e}{2}\left( \alpha_2(ik_x)^{*}\right),\\ F_z&=&|{E_0}|^{2}\frac{\Re
e}{2}\left( \alpha_2(ik_z)^{*}\right), \ee for $s$-polarization and

\be\label{forcecylindre} F_x&=&|{E_0}|^{2}\frac{\Re e}{2} \left(
\frac{2z_0^2\alpha_1(ik_x)^*} {2z_0^2+\alpha_1\Delta} \right),\\
\label{cylindrefz}
F_z&=&|{E_0}|^{2}\frac{\Re e}{2}\left(\frac{2z_0^2\alpha_1(ik_z)^*}
{2z_0^2+\alpha_1\Delta} +\frac{2z_0|\alpha_1|^2\Delta}
{|2z_0^2+\alpha_1\Delta|^2}\right),\ee for $p$-polarization.
$|{E_0}|^{2}=|{E_0}_y|^{2}$ for $s$-polarization and
$|{E_0}|^{2}=|{E_0}_x|^{2}+|{E_0}_z|^{2}$ for $p$-polarization. If,
again, $z_0\rightarrow\infty$ or $\Delta=0$, there is no interaction
between the cylinder and the surface, then we find the same equations
as those established for the sphere with only a replacement of
$\alpha$ by $\alpha_1$ or $\alpha_2$, depending on the polarization.
Concerning the force along the $X$-direction, we have the same effect
as for the sphere, namely $F_x$ has the sign of $k_x$.

\section{Numerical results and discussion}\label{result}

In this section we present numerical results on forces acting on
either a small sphere or a small cylinder. Theses forces are
normalized in the form $F_u/|E_0|^{2}$ where $F_u$ is the
$u$-component of the force, and $|E_0|$ stands for the modulus of the
incident field at the center of either the sphere or the cylinder. All
calculations are done for a body in glass ($\varepsilon=2.25$), at a
wavelength of 632.8nm, in front of a flat surface
($\epsilon=\varepsilon=2.25$) illuminated from the glass side by
internal reflection (Fig.~1).

\subsection{Results for a small sphere}\label{resultsphere}

We have first checked our CDM calculation by comparing it with the
well known Mie scattering results for a sphere in free space
illuminated by a plane wave.~\cite{hulst} The force
is:\be\label{forcemie}
\ve{F}_{Mie}=\frac{1}{8\pi}|E_0|^{2}\left(C_{ext}-\overline{\cos
\theta}C_{sca}\right)\frac{\ve{k}_0}{k_0} \ee where $C_{ext}$ denotes
the extinction cross section, $C_{sca}$ the scattering cross section,
and $\overline{\cos \theta}$ the average of the cosine of the
scattering angle. Calculations are done for a sphere of radius
$a=10$nm.}
\end{multicols}

\begin{table}[h]
\begin{center}
\begin{tabular}{|c|c|c||c|c|c||c|c||c|}
%\hline
\multicolumn{3}{|c||}{CDM-A} &\multicolumn{3}{|c||}{CDM-B}&\multicolumn{2}{|c||}{dip. app.} & Mie \\
\hline
force & $N$ & \%(Mie) & force & $N$ & \%(Mie) & force & \%(Mie) & force\\
\hline
2.8119E-22 & 81 & 0.46  &  2.8338E-22 & 81 &1.24 & 2.8027E-22 & 0.13 &2.7991E-22 \\
\cline{1-6}
2.8181E-22 & 912 & 0.68 & 2.8243E-22  & 912 & 0.91 &  &  &\\
\cline{1-6}
2.8151E-22 & 1791 & 0.57 & 2.8194E-22  & 1791 & 0.73  &  &  &\\
\cline{1-6}
2.8151E-22 & 2553 & 0.57 & 2.8186E-22  & 2553 & 0.70 &  &  & \\
%\hline
\end{tabular} 
\caption{Force on a sphere of radius $a=10$nm in free space. Numerical
results for different number of subunits $N$ in the CDM-A,
CDM-B. Comparison of calculation with the dipolar approximation and
Mie's calculation. \%(Mie) is the relative difference (in percent)
between the exact Mie calculation and the method
used.\label{tasphere}}
\end{center}  
\end{table}

\begin{multicols}{2}
Table~\ref{tasphere} compares the force obtained from the CDM on
using, without any approximation, either the method developed in
Section~\ref{CDMA} (CDM-A) or that from Section~\ref{CDMB} (CDM-B),
and from the dipolar approximation (dip. app.) presented in
Section~\ref{dipapps}, with the Mie calculation (\%(Mie) is the
relative difference in percent between the Mie result and the other
corresponding method). For an incident field with $|E_0|=94825$V/m,
which corresponds to a power of $1.19$mW distributed on a surface of
$10\mu$m$^2$, the force on the sphere in MKSA units is $2.7991\times
10^{-22}$ Newtons. One can see that for both CDM-A and B the
convergence is reached even for a coarse discretization, and, hence,
either one of the two CDM approaches can be used. As regards the
dipolar approximation, we conclude that it is perfectly valid to use
it for a sphere of radius $a=$10nm ($a/\lambda<0.016$).

Now that we have validated our methods (both analytic and CDM) we
proceed to take the surface into account. It should be remarked that
with the CMD-A it is not possible to compute the force when the sphere
is on the surface.  This is because for an observation point very
close to the sphere, the electromagnetic field values are affected by
the discretization of the sphere, and so the field is not correctly
computed. An empirical criterion that we have found~\cite{chaumet} is
that the electric field must be computed at least at a distance $d$
from the sphere but this criterion depends on the relative
permittivity. For more precision about the dependence of the criterion
and the relative permittivity one can look to
Ref.~[\ref{draine}]. With the CDM-B this problem does not occur
because with this approach it is not necessary to obtain the field
outside the sphere.

\begin{figure}[H]
\begin{center}
\includegraphics*[draft=false,width=110mm]{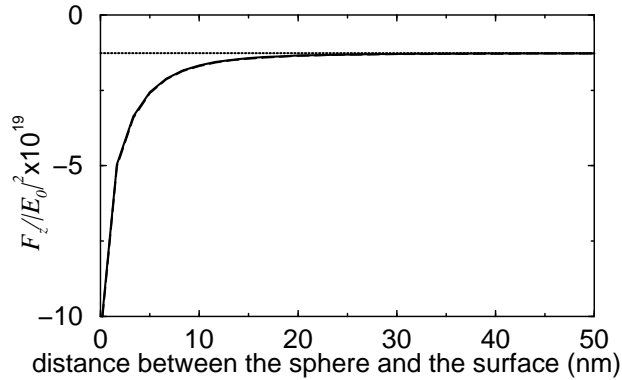}
\end{center}
\caption{Normalized force in the $Z$-direction on the sphere of
$a=10$nm versus distance $Z$. The angle of incidence of illumination
is $\theta=42^{\circ}$ in $p$-polarization. The full line represents
the exact calculation with CDM-B, the dashed line corresponds to the
static approximation with CDM-B, and the dotted line is the
calculation without interaction between the sphere and the surface.}
\end{figure}

In all figures shown next, we plot the force versus the distance $z$
between the sphere (or the cylinder, see Section~\ref{resultcyl}) and
the plane, (notice that we represent by $z_0$ the distance between the
center of the sphere, or cylinder, and the plane). The calculation
using the dipole approximation, as well as the CDM(A-B), has been done
with the static approximation for the field susceptibility of the
surface (SAFSS). However, the distance between the sphere and the
surface goes generally up to 100nm. In order to justify the study of
the force at distances about 100nm between the sphere and the plane
through a calculation done in the static approximation, we plot in
Fig.~2 the normalized force $F_z$ for $p$-polarization, with a sphere
of radius $a=10$nm, at an angle of incidence $\theta=42^{\circ}$,
without any approximation with the CDM-A (namely, taking into account
all retardation effects) with the SAFSS, and with the approximation in
which no interaction between the sphere and the surface is
considered. The difference between SAFSS and the exact calculation is
less than 1.5\%. This is in fact logical. Near the surface, the SAFSS
is correct, far from the surface, however, the field susceptibility
associated with the surface in the exact calculation is significantly
different from the field susceptibility derived from a static
approximation. Nevertheless, for distances larger than $z=30$nm the
curves overlap because the sphere does not feel the substrate at this
distance. This is manifested by a difference of only 2\% between the
exact calculation result and that computed without addressing the
surface (horizontal line).

\begin{figure}[H]
\begin{center}
\resizebox{100mm}{!}{\input{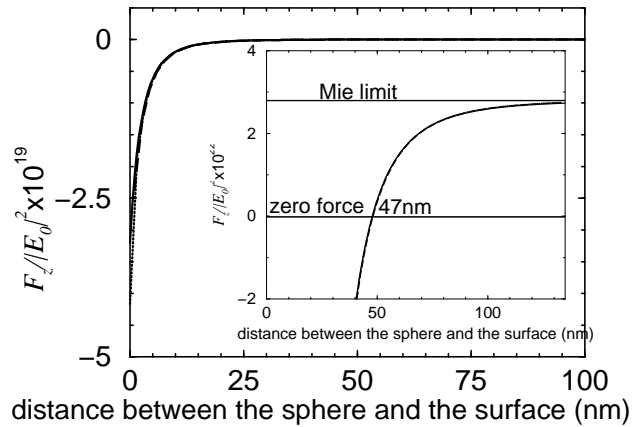}}
\end{center}
\caption{Normalized force in the $Z$-direction on a sphere of radius
$a=10$nm. The full line corresponds to the dipole approximation, the
dashed line to the CDM-A, and the dotted line to the CDM-B. The angle
of incidence is $\theta=0^{\circ}$. The inset shows the force near
$z=50$nm. We show the zero force and the force computed from Mie's
limit with Eq.~(\ref{forcemie}).}
\end{figure}

Fig.~3 shows the normalized force for light at an angle of incidence
$\theta=0^{\circ}$. The curves corresponding to CDM-A and B are
similar, and the dipole approximation appears slightly above when the
sphere is close to the surface. This may seem strange at first sight
in view of the good results presented in Table~\ref{tasphere} (we will
discuss it later). We can see that although the illuminating wave is
propagating, if the sphere is near the surface, it is attracted
towards it, opposite to the propagation direction. To understand this,
we look at Eq.~(\ref{force3dsb}), established with the dipole
approximation with the values $k_x=0$, $k_z=k_0$, ${E_0}_z=0$ which
corresponds to $\theta=0^{\circ}$. After some approximations (namely
$(4/9)k_0^6\alpha_0^2\ll 1$ which implies
$|\alpha|^2\simeq\alpha_0^2$), the force can be written:
\be\label{forceexpl}
F_z=\frac{|E_0|^{2}64z_0^6}{|8z_0^3+\alpha\Delta|^2}
\left(\alpha_0^{2}k_0^{4}/3+ \frac{3\alpha_0^{2}\Delta}{32
z_0^4}\right).\ee The factor before the bracket of
Eq.~(\ref{forceexpl}) corresponds to the intensity of the field at the
position of the sphere. The first term in the bracket of this equation
is due to the light scattering on the particle (as in free space) and
is always positive. The second term in the bracket is always negative
as $\Delta<0$. Therefore, the relative weight of the two terms in
Eq.~(\ref{forceexpl}) determines the direction of $F_z$. $F_z$ given
by Eq.~(\ref{forceexpl}), becomes zero for: \be
\label{zeroforces}z_0^4=\frac{9(\varepsilon-1)}{32
k_0^4(\varepsilon+1)}.  \ee Hence, in our example we find
$z_0=57$nm. Below the value of Eq.~(\ref{zeroforces}) the force is
attractive towards the surface, and above this value the sphere is
pushed away. This is seen in the inset of Fig.~3 which enlarges those
details. We find $F_z=0$ at $z=47$nm namely at $z_0=$(47+10)nm=57nm,
which is exactly the same value previously found. Physically, the
attraction of the sphere is due to the second term of
Eq.~(\ref{forceexpl}) which corresponds to the interaction of the
dipole with its own evanescent field reflected by the surface. Now we
can explain the discrepancy between the dipole approximation and the
CDM as regards the good results obtained in free space. In fact, when
the computation is done in free space the field can be considered
uniform over a range of $20$nm. However, in an evanescent field, the
applied field is not uniform inside the sphere and the Clausius
Mossotti relation is less adequate, hence, the dipole approximation
departs more from the exact calculation. However, when the sphere is
out from the near field zone, the three methods match together (see
the inset of Fig.~3). We can also see in the inset of Fig.~3, that
these three curves tend towards the Mie limit because at large
distance there is no interaction with the surface.

\begin{figure}[H]
\begin{center}
\includegraphics*[draft=false,width=90mm]{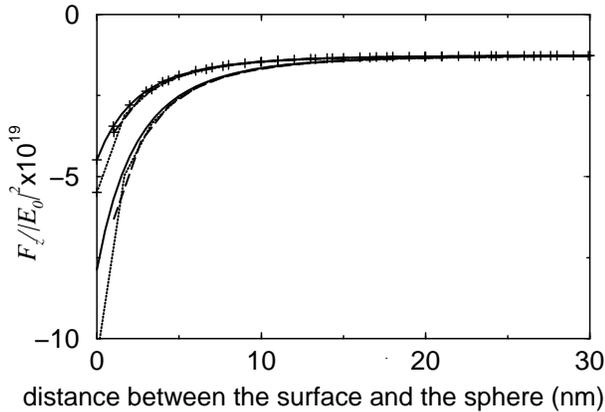}
\end{center}
\caption{Normalized force in the $Z$-direction acting on the sphere
with $a=$10nm. The angle of incidence $\theta=42^{\circ}$ is larger
than the critical angle $\theta_c=41.8^{\circ}$. The full line
corresponds to the dipole approximation, the dashed line to the CDM-A,
and the dotted line to the CDM-B. Curves without symbols are for
$p$-polarization, and those with symbol + are for $s$-polarization.}
\end{figure}

Fig.~4 shows the $z$-component of the normalized force when the
incident wave illuminated at
$\theta=42^{\circ}>41.8^{\circ}=\theta_c$. Then, for $s$-polarization
we can write Eq.~(\ref{force3dsb}) as: \be\label{force3dsbeva}
F_z=\frac{|{E_0}_y|^{2}}{|8z_0^{3}+\alpha\Delta|^{2}}\left [
-4z_0^3\gamma\alpha_0(\alpha_0\Delta+8z_0^3)+6z_0^2\alpha_0^2\Delta\right]
.\ee It is easy to see that for a dielectric sphere both the first and
second terms within the brackets of Eq.~(\ref{force3dsbeva}) are
always negative. Hence, the sphere is always attracted towards the
surface (the same reasoning can be done for $p$-polarization). Near
the surface the force becomes larger because of the interaction of the
sphere with its own evanescent field. We notice that the normalized
force becomes constant at larger $z$.  This constant reflects the fact
that the force decreases as $e^{-2\gamma z}$ from the surface.

\subsection{Results for a small cylinder}\label{resultcyl}

Let us now address an infinite cylinder. Since the CDM method used
here works in three dimensions, we have computed the force on a finite
length cylinder. In order to verify this approximation, we once again
compare the force, obtained in free space from the CDM with different
cylinders lengths, with that from a calculation done with the dipole
approximation established in Section~\ref{dipappc}, and that from an
exact calculation for an infinite cylinder~\cite{hulst} (i.e. the well
known 2-$D$ version for cylinders of the Mie calculation for spheres,
hence referred to in the table as ``Mie'') . We consider a radius of
the cylinder, $a=10$nm, with the same spacing lattice as for the case
of the sphere, namely 81 subunits. We have seen that this value of
$d=4$nm gives consistent results. In all cases we compute the force
per unit length of the cylinder.

The first case addressed is with the electric field perpendicular to
the axis of the cylinder ($p$-polarization). The results are given in
Table~\ref{tacyindrep}. The second case considered ($s$ polarization)
in Table~\ref{tacyindres}.

We notice that the dipole approximation gives correct results for
$p$-polarization, but it is worse for $s$-polarization. If we compare
the CDM-A and CDM-B, we see that they both give the same results.  But
we also see that the length of the cylinder has a great influence on
them, although up to a different extent according to whether we deal
with $p$ or $s$-polarization. For $p$-polarization, the simulation of
an infinite cylinder becomes correct at $L\simeq\lambda/2$ and for
$s$-polarization only it is so at $L\simeq2\lambda$.  This can be
understood by the fact that in $p$-polarization the electric field is
continuous at the end of the cylinder, thus, the end has not a large
influence on the field computed around the cylinder. However, in
$s$-polarization the field is discontinuous at the end of the cylinder
and then the field will strongly vary around this end and so will do
the force. This is why in $s$-polarization it is necessary to consider
cylinders with large lengths in order to avoid edge effects. Now let
us address the presence of the plane surface to compute the force. We
consider the cylinder length $L=1551$nm. Like for the sphere, we
address both $\theta=0^{\circ}$, (Fig.~5), and $42^{\circ}$, (Fig.~6).
The curves from CDM-B stop at $z=$10nm due to the disadvantage
previously remarked.
\end{multicols}

\begin{table}[h]
\begin{center}
\begin{tabular}{|c|c|c||c|c|c||c|c||c|}
%\hline
\multicolumn{3}{|c||}{CDM-A} &\multicolumn{3}{|c||}{CDM-B}&\multicolumn{2}{|c||}{dip. app.} & Mie \\
\hline
force & L(nm) & \%(Mie) & force & L(nm) & \%(Mie) & force & \%(Mie) & force\\
\hline
2.1540E-13 &197 & 24 &2.1625E-13  & 197  &24 & 2.8433E-13   & 0.27 & 2.8354E-13  \\
\cline{1-6}
2.9906E-13  & 391 & 5.47 &3.0013E-13  & 391  &5.85 &  &  &       \\
\cline{1-6}
2.7907E-13  & 777 & 1.58 &2.8000E-13  & 777  &1.25 &  &  &       \\
\cline{1-6}
2.8661E-13  & 1164 & 1.08  &2.8756E-13  & 1164  & 1.42 &  &  &       \\
\cline{1-6}
2.8347E-13   & 1551 & 0.03 &2.8439E-13  & 1551 & 0.30 &   &  &       \\
%\hline
\end{tabular} 
\caption{Force on a finite cylinder of radius $a=10$nm in free
space. The discretization interval is $d=4nm$. Numerical results are
presented for different lengths $L$ of the cylinder for both CDM-A and
CDM-B. Comparison is made with both the dipolar approximation and
Mie's calculation. \%(Mie) is the relative difference between the
exact Mie calculation for an infinite cylinder and the method
used. Calculations are done for the field perpendicular to the axis of
the cylinder.\label{tacyindrep}}
\end{center}  
\end{table}

The second case considered is with the electric field parallel to the
axis of the cylinder ($s$-polarization):
\begin{table}
\begin{center}
\begin{tabular}{|c|c|c||c|c|c||c|c||c|}
%\hline
\multicolumn{3}{|c||}{CDM-A} &\multicolumn{3}{|c||}{CDM-B}&\multicolumn{2}{|c||}{dip. app.} & Mie \\
\hline
force & L(nm) & \%(Mie) & force & L(nm) & \%(Mie) & force & \%(Mie) & force\\
\hline
 0.5649E-12 &197 & 63 &0.2163E-12  & 197 & 86 &  1.5015E-12  & 2.31 &  1.5370E-12 \\
\cline{1-6}
 0.9986E-12  & 391 & 35 &1.0021E-12  & 391  & 35 &  &  &       \\
\cline{1-6}
 1.3059E-12 & 777 & 15.0 &1.3103E-12  & 777  & 14.7 &  &  &       \\
\cline{1-6}
1.3971E-12  & 1164 & 9.10 &1.4018E-12  & 1164  & 8.80 &  &  &       \\
\cline{1-6}
1.4430E-12   & 1551 & 6.12  &1.4479E-12  & 1551 & 5.80 &  &  &       \\
%\hline
\end{tabular} 
\caption{The same as in Table~\ref{tacyindrep} but for the electric
field parallel to the axis of the cylinder.\label{tacyindres}}
\end{center}  
\end{table}

\begin{multicols}{2}

\begin{figure}[H]
\begin{center}
\includegraphics*[draft=false,width=90mm]{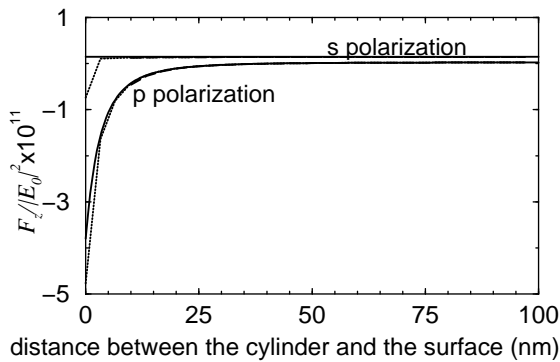}
\end{center}
\caption{Normalized force in the $Z$-direction on a cylinder with
radius $a=$10nm, $\lambda=$632.8nm, and $\varepsilon=2.25$. The light
angle of incidence is $\theta=0^{\circ}$. The full line corresponds to
the dipole approximation, the dashed line to the CDM-A, and the dotted
line to the CDM-B.}
\end{figure}

Concerning Fig.~5, if we focus on $F_z$ for $p$-polarization, we can
write this force approximated from Eq.~(\ref{cylindrefz}) by:
\be\label{cylindrefz1}
F_z=\frac{4z_0^4|E_0|^2}{|2z_0^2+\alpha_1\Delta|^2} \left(
(\alpha_1^0)^2k_0^3\pi/4+\frac{(\alpha_1^0)^2\Delta}{4z_0^3}\right).\ee
Eq.~(\ref{cylindrefz1}) is of the same form as
Eq.~(\ref{forceexpl}). Hence, the same consequence is derived: near
the surface the cylinder is attracted towards the plane surface. But
far from the plane the cylinder is pushed away because at this
distance the cylinder cannot interact with itself. Like for the
sphere, we can compute the distance $z_0$ at which the force is null:
\be z_0^3=\frac{(\varepsilon-1)}{\pi k_0^3(\varepsilon+1)}, \ee which
in our illustration leads to $z_0=50$nm. Although we do not present
now an enlargement with details of Fig.~5, we have found the value
$z_0$=(40+10)nm=50nm. The cylinder in $p$-polarization has the same
behaviour as the sphere. However, in $s$-polarization there is a
difference. Then the force obtained from the dipolar approximation is
always constant because there is no interaction with the surface. This
is clear from Eq.~(\ref{forcecylindre}), and it is due to the fact
that in the electrostatic limit the field susceptibility $S_{yy}$
tends to zero, and then there is no influence of the surface on the
cylinder. This is a consequence of the continuity of both the field
and its derivative of both the plane and the cylinder.~\cite{jjg2}
Therefore, the cylinder does not feel the presence of the plane, and
as the wave is propagating, the force is positive thus pushing the
cylinder away from the plane with magnitude values given by
table~\ref{tacyindres}. Notice that the force obtained from CDM-B,
when the sphere is in contact with the surface, becomes negative in
$s$-polarization. This is due to the diffraction of the field at the
end of the cylinder, which induces a component perpendicular to the
plane, and therefore an attractive force.

\begin{figure}[H]
\begin{center}
\includegraphics*[draft=false,width=80mm]{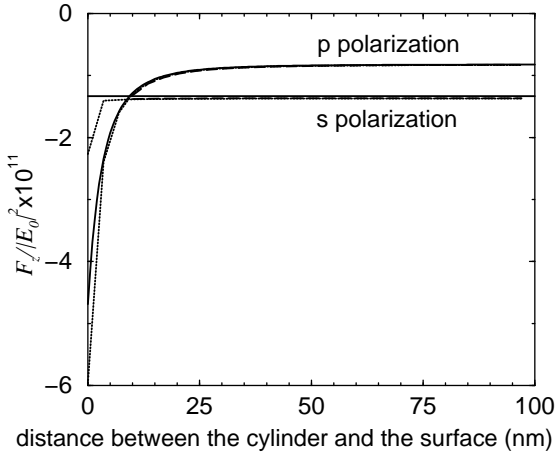}
\end{center}
\caption{Normalized force in the $Z$-direction on the same cylinder as
describes in Fig.~5 but with an angle of incidence $\theta=42^{\circ}$
larger than the critical angle $\theta_c=41.8^{\circ}$.  The full line
corresponds to the dipole approximation, the dashed line to the CDM-A,
and the dotted line to the CDM-B.}
\end{figure}

In the case represented in Fig.~6, like for the sphere, we observe a
force always attractive ($F_z<0$) whatever the polarization. For
$p$-polarization we have exactly the same behaviour as for the
sphere. However, for $s$-polarization the normalized force is always
constant whatever the distance between the cylinder and the surface,
due to the same reason as before, namely, $S_{yy}=0$. Only when the
cylinder is on the surface, we can see from the CDM-B calculation that
the force is slightly more attractive for the same reason previously
quoted.

\subsection{Results for a sphere beyond the Rayleigh regime}\label{bigsphere}

Let us now consider a sphere of radius $a=$100nm. This size is far
from the Rayleigh scattering regime ($\approx\lambda/3$). As in
previous cases, we first validate our method with the aid of Mie's
calculation in free space. The following table shows the results:

As before, as $d$ decreases, the CDM results tend to the Mie's
calculation. The error never exceeding 1.7\%. Now, we address the
presence of a flat dielectric surface. The forces, to be shown next,
are computed with CDM-B only since the particle can be in contact with
the surface.

\begin{figure}[H]
\begin{center}
\resizebox{80mm}{!}{\input{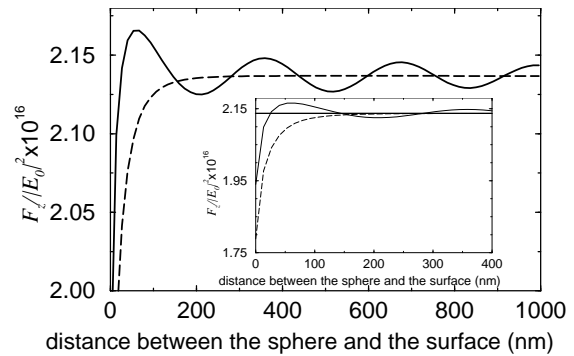}}
\end{center}
\caption{Normalized force in the $Z$-direction on a sphere with radius
$a=$100nm, $\lambda=$632.8nm, and $\varepsilon=2.25$. The light angle
of incidence is $\theta=0^{\circ}$. The full line corresponds to the
exact calculation with CDM-B, and the dashed line represents the
static approximation.}
\end{figure}

\end{multicols}

\begin{table}
\begin{center}
\begin{tabular}{|c|c|c||c|c|c||c|}
\hline
\multicolumn{3}{|c||}{CDM-A} &\multicolumn{3}{|c||}{CDM-B}& Mie \\
\hline
force & $N$ ($d=$nm) & \%(Mie) & force & $N$ ($d=$nm) & \%(Mie)  & force\\
\hline
2.1355E-16 & 280 (25) & 1.31  &  2.1439E-16 & 280 (25) &1.71 & 2.1080E-16 \\
\cline{1-6}
2.1353E-16 & 912 (17) & 1.30 & 2.1402E-16  & 912 (17) & 1.53   &\\
\cline{1-6}
2.1332E-16 & 1791 (13) & 1.20 & 2.1367E-16  & 1791 (13) & 1.37   &\\
\cline{1-6}
2.1312E-16 & 4224 (10) & 1.11 & 2.1333E-16  & 4224 (10) & 1.21  & \\
\hline
\end{tabular} 
\caption{Force on a sphere of radius $a=100$nm in free
space. Numerical results are for different number of subunits $N$ in
the CDM-A, CDM-B. Comparison with Mie's
calculation.\label{tabigsphere}}
\end{center}  
\end{table}

\begin{multicols}{2}

In Fig.~7 we present the case $\theta=0^{\circ}$. We have plotted two
curves: the exact calculation and the SAFSS done with $N=1791$. In the
inset of Fig.~7, we see that even near the surface the SAFSS is not
good. This is due to the large radius of the sphere, then the
discretization subunits on the top of the sphere are at 100nm from the
surface, and thus the effects of retardation are now important. The
SAFSS calculation also shows that at a distance of 200nm (which
corresponds to the size of the sphere: $2a$=200nm) the sphere does not
feel the surface, as manifested by the fact than then the curve
obtained from this computation reaches the Mie scattering limit
previously obtained in Table~\ref{tabigsphere} (cf. the full
horizontal line in the inset). Hence, we conclude that evanescent
waves are absent from the interaction process at distances beyond this
limit. From the exact calculation we obtain a very low force near the
surface, due to the interaction of the sphere with itself. This effect
vanishes beyond $z\approx 50$nm where oscillations of the force $F_z$
take place with period $\lambda/2$. As these oscillations do not occur
in the SAFSS, this means that they are due to interferences from
multiple reflections between the surface and the sphere. As expected,
they decrease as the sphere goes far from the surface.

\begin{figure}[H]
\begin{center}
\resizebox{80mm}{!}{\input{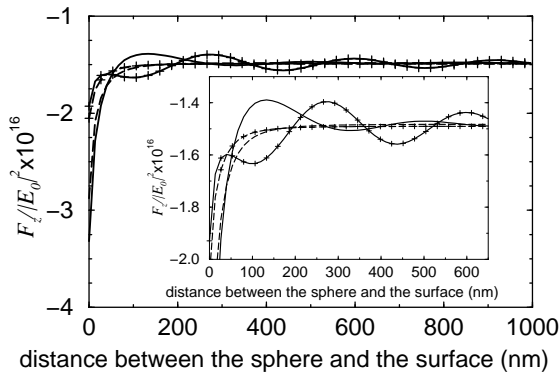}}
\end{center}
\caption{Normalized force in the $Z$-direction on a sphere with radius
$a=$100nm, $\lambda=$632.8nm, and $\varepsilon=2.25$. The light angle
of incidence is $\theta=42^{\circ}>\theta_c$. The full line
corresponds to the exact calculation with CDM-B, and the dashed line
to the static approximation. the curves without symbol are in
$p$-polarization, and those with the $+$ symbol in $s$-polarization.}
\end{figure}

Fig.~8 shows the force computed with an angle of incidence
$\theta=42^{\circ}$. We plot the exact calculation (full line) and the
SAFSS (dashed line) both for $p$-polarization (no symbol) and
$s$-polarization ($+$ symbol). Once again, we see that the SAFSS is
not adequate even near the surface. On the other hand, in the exact
calculation, the two polarizations show oscillations of the force
$F_z$ with period $\lambda/2$. However, there is a large difference of
magnitude of these oscillations between the two polarizations (see
inset of Fig.~8). To understand this difference, we must recall that
the sphere is a set of dipoles. When a dipole is along the
$Z$-direction there is no propagating wave in this direction. But if
the dipole is oriented in the $X$(or $Y$)-direction, its radiation is
maximum in the $Z$-direction. However, in $s$-polarization all dipoles
are, approximately, parallel to the surface, so there is an important
radiation from the dipole in the $Z$-direction, and, consequently,
between the sphere and the surface.

\section{conclusions}

In this paper we have presented exact three dimensional calculations
based on the Coupled Dipole Method and an analytical expression for
the force on either a sphere or an infinite cylinder, both in front of
a flat dielectric surface. The results for small bodies show that,
whatever the polarization, in the case of a sphere, and in
$p$-polarization for the cylinder, the force always has the same
behaviour: namely, in the case of illumination under total internal
reflection, the particle is always attracted towards the surface. A
surprising result in the case when the illuminating beam is
perpendicular to the surface and the object remains sticked to the
surface, is that then the force is attractive due the interaction of
the particle with itself, and therefore this object keeps sticked to
the surface. However, when the object is far from the surface, the
force becomes repulsive, as one would have expected.

For $s$-polarization, the cylinder does not ``feel'' the presence of
the substrate. This is more noticeable for a propagating wave, namely,
at angles of incidence lower than the critical angle. However, when an
evanescent wave is created by total internal reflection, the force is
attractive under $s$-polarization.

The scope of the static calculation for this configuration has been
validated. We have also shown the advantage of having an analytical
form which shows the contribution of the incident field on the
particle, as well as that of the force induced by the sphere (or
cylinder) on itself, thus yielding a better understanding of the
physical process involved.

For bigger spheres, we have observed somewhat different effects of the
forces. Under the action of evanescent waves, the force is always
attractive, but it always becomes repulsive when it is due to
propagating waves. Unlike the case of small sphere, there is no 
point of zero force.

\section{Acknowledgments}

This work has been supported by the European Union, grant ERBFMRXCT
98-0242 and by the DGICYT, grant PB 98-0464.

%%%%%%%%%%%%%%%%%%%%%%%%%%%REFERENCES%%%%%%%%%%%%%%%%%%%%%%%%%%%%

\end{multicols}


\begin{references}
\bibitem{force} In fact $\ve{F}$ is the time averaged force on the
particle.
\end{references}
\end{document}